\documentclass[a4paper]{jpconf}
\usepackage{graphicx}
\def\yohkoh{{\sl Yohkoh}}
\def\hinode{{\sl Hinode}}
\def\sdo{{\sl SDO}}

\newcommand{\ltsim}{\raisebox{-1.0ex}{$\stackrel{\textstyle<}{\sim}$}}

\begin{document}
\title{Coronal Jets, and the Jet-CME Connection}

\author{Alphonse C. Sterling}

\address{NASA/Marshall Space Flight Center, Huntsville, AL, 35806, USA}

\ead{alphonse.sterling@nasa.gov}

\begin{abstract}

Solar coronal jets have been observed in detail since the early 1990s.
While it is clear that these jets are magnetically driven, the details of
the driving process has recently been updated. Previously it was suspected that the jets were a consequence of
magnetic flux emergence interacting with ambient coronal field. New evidence however indicates that
often the direct driver of the jets is erupting field, often carrying cool material
(a ``minifilament''), that undergoes interchange magnetic reconnection with preexisting field (\cite{sterling.et15}). More recent work indicates that the trigger for eruption of the minifilament is
frequently cancelation of photospheric magnetic fields at the base of the minifilament. These erupting
minifilaments are analogous to the better-known larger-scale filament eruptions that produce solar
flares and, frequently, coronal mass ejections (CMEs). A subset of coronal jets drive narrow
``white-light jets,'' which are very narrow CME-like features, and apparently a few jets can 
drive wider, although relatively weak, ``streamer-puff'' CMEs.  Here we summarize these recent findings.

\end{abstract}

\section{Introduction}

Coronal jets are transient, long and relatively narrow features seen in EUV and X-ray images that shoot out from the solar
surface and into the corona.  While there were a smattering of earlier observations \cite{raouafi.et16},
jets were first studied in great detail with the Soft X-ray Telescope (SXT) on the {\sl Yohkoh} spacecraft,
launched in 1991.  Because it was relatively sensitive to comparatively hot coronal emissions, the jets that appeared in the
vicinity of active regions tended to be easiest to detect \cite{shibata.et92,shimojo.et96}. 
SXT observations found jets to reach lengths of $1.5\times 10^4$~km \cite{shimojo.et96}, and
temperatures to range over 3~MK---8~MK with an average of 5.6~MK (\cite{shimojo.et00}).

Launched in 2006, the \hinode\ satellite carries the X-Ray Telescope (XRT), which is sensitive to
comparatively  cooler coronal plasmas than was SXT; in particular, it is sensitive to emissions of
1~MK---2~MK, a characteristic temperature  range for many coronal hole jets
(\cite{pucci.et13,paraschiv.et15}).  XRT found jets to be plentiful in polar coronal holes
\cite{cirtain.et07}; according to \cite{savcheva.et07}, jets occur at a rate of  about 60 day$^{-1}$ in the
two polar coronal holes, have lifetimes of $\sim$10~min, lengths of 50{,}000~km,  and widths of 8{,}000~km.

A jet consists of a spire, which extends into the corona, and a base region that is generally much brighter
than the spire.  Moreover, the brightest part of the base is often positioned asymmetrically with the spire,
being located off to one side of the base region.  These jet-base brightenings (hereafter, JBP, following \cite{sterling.et15}) have long been suspected as holding a clue to the cause of jets (\cite{shibata.et92};
also see \cite{heyvaerts.et77}).

\section{The Cause of Coronal Jets}

It was quickly realized that the energy source for jets was likely magnetic.  Magnetic field that  permeates
the solar atmosphere is fed by new magnetic flux that almost continuously ``bubbles up'' (i.e.\ emerges) 
through the  photospheric surface.  Upon reaching the surface, this emerging field spreads horizontally on
the surface  and expands outward into the higher atmosphere.  An early suggestion was that jets resulted
when flux emerging in this manner into the chromosphere and low corona runs into nearby pre-existing
magnetic field in those regions. Interactions between the two flux systems would result in magnetic
reconnection, at the interface between the emerging bipole and the pre-existing (e.g.\ open) field.  According to this emerging-flux model, this reconnection would result in a small closed magnetic 
loop field, and also new open magnetic field, where the new closed loop would form the JBP
and the jet spire plasma would flow out along newly open field.

New studies became possible with the launch of the Solar Dynamics Observatory (\sdo) satellite in  2010. Its
Atmospheric Imaging Assembly (AIA) instrument images the Sun in multiple full-disk EUV and UV wavelengths
(\cite{lemen.et12}) at 12~s cadence (in the EUV bands) and with $0.''6$ pixels, allowing for a much broader
range of coverage of jet emissions than was possible with the X-ray instruments alone (both SXT and XRT had
somewhat lower resolution than AIA, and variable cadences and fields of view).  Also critical for jet
studies is  \sdo's Helioseismic and Magnetic Imager (HMI), which takes full-disk line-of-sight magnetograms
at a cadence of 45~s, among other capabilities (\cite{scherrer.et12}).  Using these instruments and other
resources, several jets were observed on the solar disk.  They frequently showed the jets to emanate from
locations where flux {\it cancelation} was occurring, without emergence (e.g., \cite{hong.et11,
huang.et12,young.et14a,young.et14b,adams.et14}).  In some cases emergence was seen at the base locations of jets, but
usually in conjunction with flux cancelation (e.g., \cite{shen.et12, li.et15}). Some pre-\sdo\ observations
also showed emergence occurring together with cancelation at jet bases (\cite{shimojo.et98}).  Therefore it seemed that flux emergence was not essential for jet formation.

Using AIA EUV images, \cite{adams.et14} found that a jet they observed appeared to form as the  result of an
eruption of a small-scale filament, which they called a minifilament, and that the JBP was  a miniature
flare that occurred in conjunction with eruption of the minifilament.  In an extensive study of 20 randomly
selected near-limb jets, \cite{sterling.et15} verified that the minifilament picture holds generally.   They
concluded that at least many (if not all) jets are a miniature version of the familiar full-scale eruptions
that include an erupting filament, a solar flare, and often a coronal mass ejection (CME)\@.  Although their
data were too near the limb for reliable comparisons with magnetograms, they argued that, on balance, the
existing studies suggested that flux cancelation was often crucial to jet occurrence.

To investigate this further, \cite{panesar.et16b} and  \cite{panesar.et18} studied substantial numbers
($\sim$10) of on-disk jets, respectively in quiet Sun and coronal holes.  For all of those examined jets, 
they found that cancelation occurred at the base of the jets. In some of those cases, flux emergence was
also occurring, but in those cases the jet occurred at the neutral line where one pole of the emerging flux
was canceling with pre-existing opposite-polarity field; this was also found by, e.g., \cite{liu.et11} and
\cite{huang.et12}.   Thus this strongly supports that magnetic flux cancelation is the trigger for onset of
at least many jets. Moreover, for the jets of the quiet region study (\cite{panesar.et16b}), the study of
\cite{panesar.et17} found that the minifilaments that erupted to cause the jets also formed via flux
cancelation, between about two hours and two days prior to the jet occurrence (\cite{panesar.et17}).

In the case of active regions (\cite{sterling.et16,sterling.et17}), the situation is not as  clear cut. One
reason is that minifilaments are sometimes but not always visible in the lead up to the jets.  And secondly,
the evolutionary magnetic changes are very rapid in the jetting locations, often with both flux cancelations
and flux emergence occurring either concurrently or within a short time of each other.  Nonetheless, all but
one of several active region jets studied in \cite{sterling.et16} and \cite{sterling.et17} clearly occurred
at the site of flux cancelation; one event of \cite{sterling.et16} was unclear in this regard. 
\cite{sterling.et17} considers possibilities for why minifilaments are not always seen in active region
events; these reasons include that pre-eruption minifilaments might be obscured by surrounding material, or
in some cases a flux rope might erupt without a cool-material minifilament, as is the case with some
large-scale eruptions.

Figure~1 summarizes several basic aspects of the view put forth for coronal jets in recent studies by the
author and his colleagues
(\cite{adams.et14,sterling.et15,sterling.et16,sterling.et17,panesar.et16b,panesar.et17,moore.et18}).

(This team's earlier work pointed out two ``flavors'' of jets: ``standard'' and ``blowout''
(\cite{moore.et10,moore.et13}).  While the observational aspects of those jets are described appropriately in those two papers, that was prior to our findings on the connection between minifilament eruptions and jets.  Therefore we have revised our view on the cause of standard and blowout jets; the updated description is presented in \cite{sterling.et15} and \cite{moore.et18}.)

\begin{figure}
\vspace{-5cm}
\hspace*{-4cm}\includegraphics[angle=-90,width=25cm,scale=1.1]{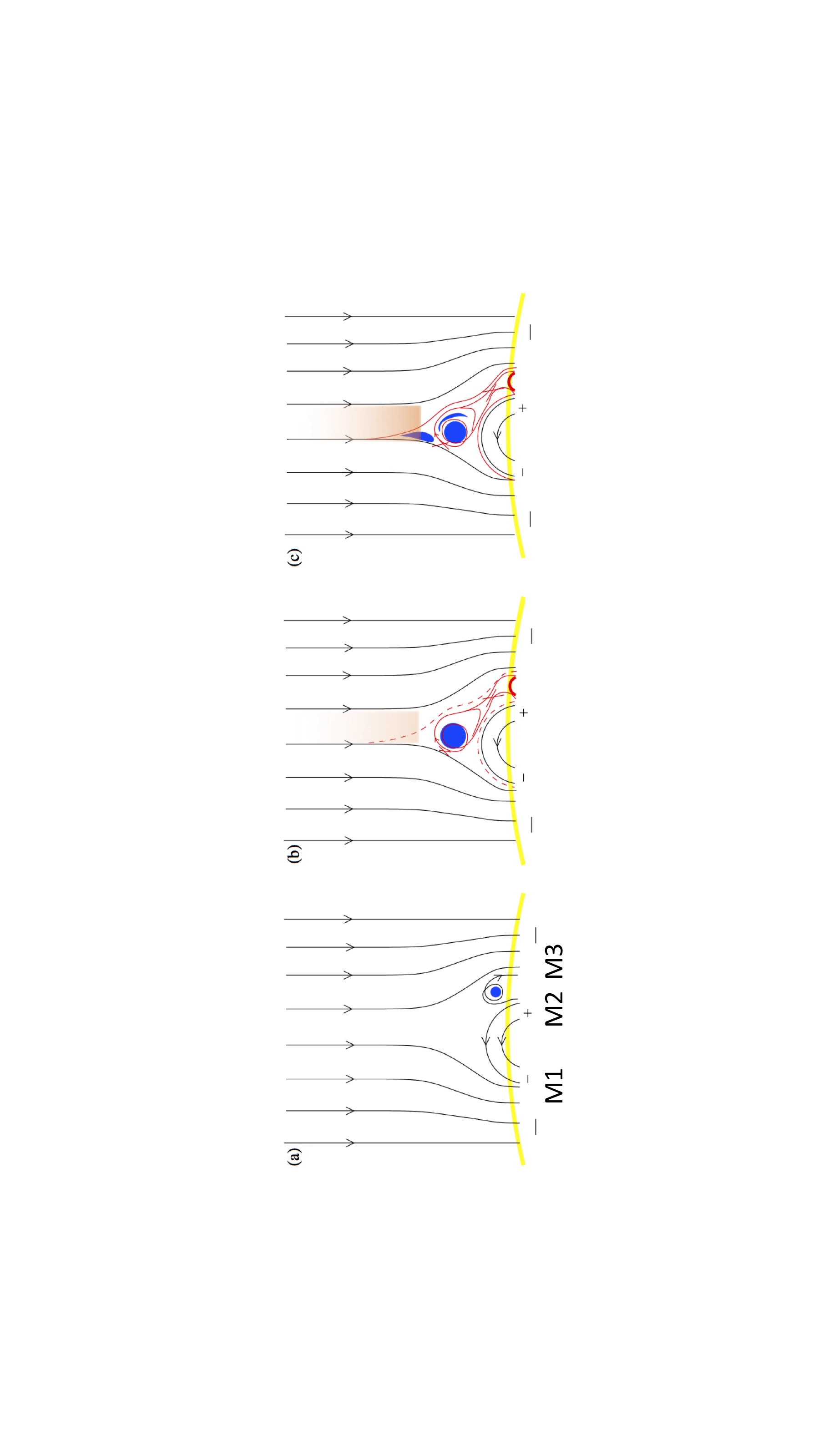}\vspace{-4cm} 
\caption{Schematic showing the {\it minifilament eruption model for coronal jets}; originally 
presented in \cite{sterling.et15}, this version is from \cite{sterling.et18}.  (a) A minifilament 
(blue disk) is in a magnetic bipole M2-M3, adjacent to a larger bipole M1-M2 (this is a 2D cut
of a 3D negative-polarity background region with an intruding positive-polarity source at M2).  The
looped field line encircling the minifilament indicates non-potential twist in the field holding the 
minifilament.  (b) The minifilament erupts, running into and undergoing interchange reconnection with opposite-polarity open (or far-reaching) field, leading to new field
lines (dotted) including new open field that guides a hot jet (shaded region), and with field below the minifilament reconnecting ``internal reconnection'' \cite{sterling.et15}) and forming the JBP 
(thick red semicircle).  (c) As the minifilament eruption progresses, eventually the interchange reconnection
erodes the minifilament field, releasing the cool minifilament material to travel outward along the jet
spire.  See \cite{sterling.et15,sterling.et16,sterling.et17,moore.et18} for more details.  This picture has been modeled numerically by \cite{wyper.et17}.}
\end{figure}

\section{Jets and CMEs}

Given the evidence that coronal jets are miniature versions of CME-producing large-scale eruptions, one
might expect that jets could also result in some version of CMEs.  Indeed ejections visible in visible-light
coronagraph images that originate from coronal jets have been detected.  These coronagraph-imaged features
seem to be of two basic types: narrow events and (somewhat) broader events, in reference to the heliocentric
angle subtended by the ejected feature.

Narrow coronagraph-imaged ejections have been called ``narrow CMEs'' by some workers
(e.g.~\cite{gilbert.et01,dobrzycka.et03,sterling.et16}), while other workers refer to them as ``white-light
jets'' (\cite{wang.et98,wang.et02}).  Depending on the study, the observed events were of width $\ltsim
10^\circ$ or $\ltsim 15^\circ$, and so in either case of much smaller angular size than that of the average
(non-halo) CME of $\sim$44$^\circ$ (\cite{gopal.et09}).  Other observations showing narrow CMEs and/or a
jet-CME connection include \cite{yashiro.et03,bemporad.et05,nistico.et09,hong.et11} and \cite{shen.et12}.

In addition to white-light-jet CMEs, somewhat broader CMEs can also result from jets
(\cite{bemporad.et05,panesar.et16a,alzate.et16}). In these cases, the jets occur in the base region of the
location that eventually erupts outward as the CME\@.  In the cited studies, the base regions of the CMEs
are bipolar active regions, and the jets occur along one edge of the active regions.  These active regions
form the bases of streamer  structures that extend far into the corona, and the ejected CMEs travel outward
along the streamers.  These CMEs tend to be weak in intensity compared to the coronagraph-bright  streamer,
and the CMEs do not totally disrupt the streamer; for these reasons we refer to  these events as ``streamer-puff'' CMEs (\cite{bemporad.et05}).  For the six streamer-puff events analyzed in \cite{panesar.et16a},
widths ranged over $25^\circ$---$50^\circ$.  Thus they are somewhat broad; broader than the above-mentioned
narrow  CMEs ($\ltsim 15^\circ$), but only reaching about the above-mentioned average width for general CMEs
($\sim$44$^\circ$).

There is evidence that both of these types of CMEs are driven by the jets, but the driving mechanism differs
between the two.  Moreover, a key component to the driving appears to be non-potential magnetic  twist
contained in minifilaments, where the twist either is added to the minifilament as it is in the process of
erupting to form the jet, or the twist exists on the minifilament prior to its eruption (in which case more
twist might be added during eruption); Figure~1 shows schematically this twist on the minifilament.

Regarding the mechanism in narrow CMEs, \cite{moore.et15} found that the jets that formed such narrow CMEs
all had a propensity to contain a relatively large amount of twist, compared to the average jet that did not
make a (narrow) CME\@.  The picture envisioned by \cite{moore.et15} for these narrow CMEs is basically that
of Figure~1, but where the reconnection in Figure~1(b) transfers a relatively large amount of twist onto the
open field (\cite{shibata.et86}).  That twist that has been added to the open field then helps propel the
jet material far enough away from the Sun so that it can show enhanced density in \sdo/LASCO C2 coronagraph
images.  

Regarding the mechanism for the streamer-puff CMEs, \cite{bemporad.et05} provided an initial suggestion. 
That study however was prior to our new understanding about jets as expressed in the minifilament eruption
model (Fig.~1).  \cite{panesar.et16a} now provides a revised view that supersedes the explanation given in
\cite{bemporad.et05}.   According to \cite{panesar.et16a}, streamer-puffs result as shown in Figure~2;
basically, the setup is as in Figure~1, but where the apparently vertical field  is actually a part of a
larger-scale loop, and that larger-scale loop forms the base region of a coronal streamer.  Upon eruption,
twist contained in the erupting minifilament is transferred to the larger-scale loop. That injected twist
drives the larger-scale loop outward along the streamer, forming the streamer-puff CME\@.

\begin{figure}
\vspace{-1cm}
\hspace*{3.5cm}\includegraphics[angle=-90,width=10cm]{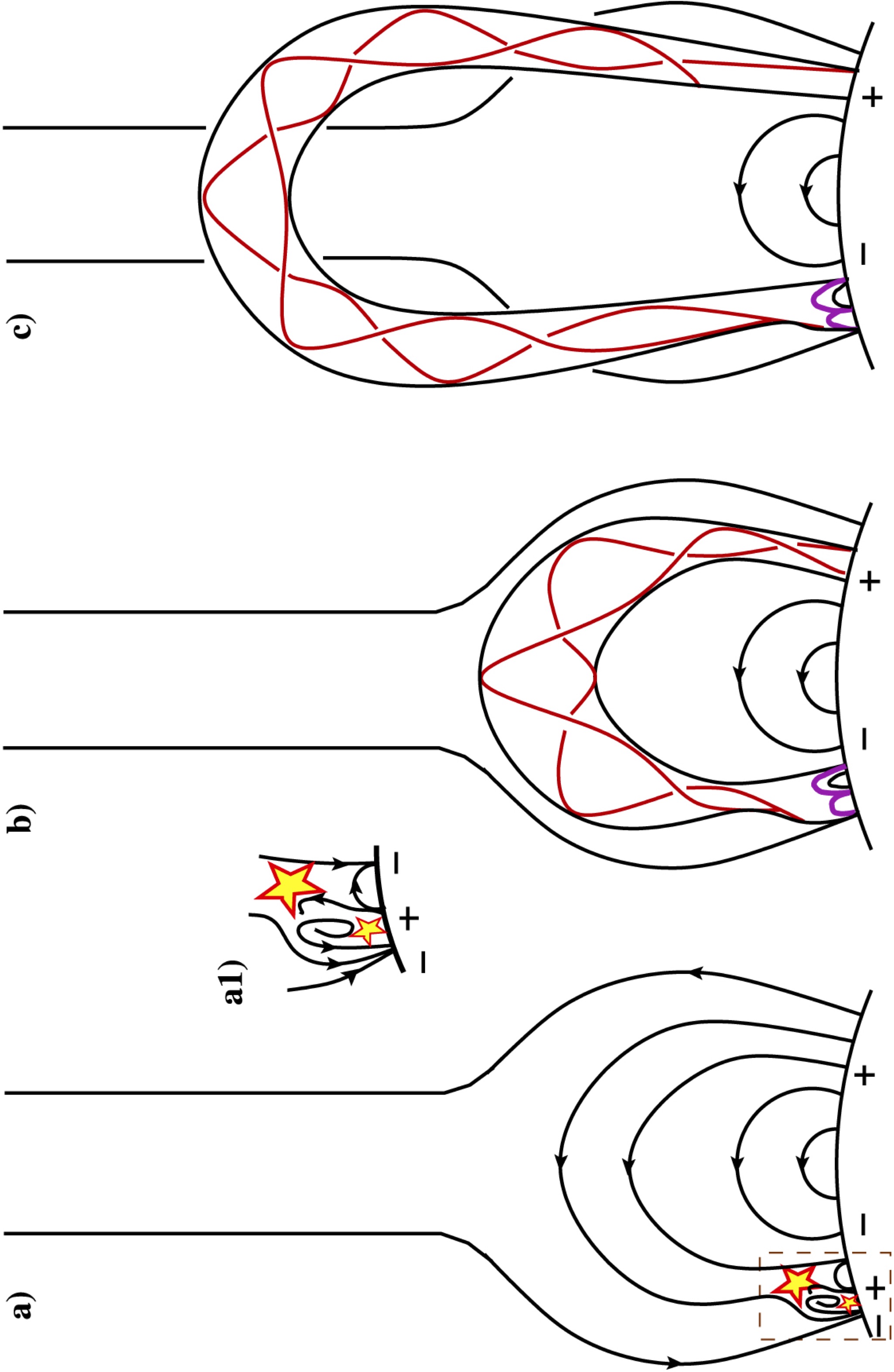}\vspace{1cm} 
\caption{Jet eruption driving a {\it streamer-puff} CME, from \cite{panesar.et16a}.  
(a) A minifilament-field (with or without a cool minifilament) erupts as in Figs.~1(a) and~1(b)
(see details in insert a1). In this case the minifilament (with a twisted-field magnetic envelope) 
erupts inside a larger-loop field (in Fig.~1(a), 
the corresponding field was either open or far-reaching; here is is ``far reaching,'' extending to the
``+'' on to the right of the erupting minifilament system).  This larger loop is part of a loop system 
that forms the base of a surrounding and overlying streamer field.  (b) Twist from the 
erupting-minifilament field is transferred
to the larger loop.  (c) This renders the larger loop unstable, causing it to erupt along the streamer to
form the streamer-puff CME\@.}
\end{figure}

\section{In Closing}

Much work has been done to advance our understanding of coronal jets over the past two decades since their
(re)discovery with the \yohkoh\ satellite.  Paper \cite{raouafi.et16} reviews much of this progress. But
that review was largely written during 2015, just as an understanding of a connection between  minifilament
eruptions and jets (\cite{sterling.et15}) was coming to the fore.  Thus, in the author's view,
\cite{raouafi.et16} under-emphasizes the importance of minifilament eruptions in jet production. Of course
more studies are required to establish with certainty (or to challenge) this idea.  Similarly, it
now appears that the importance of flux cancelation in triggering jets
\cite{young.et14a,young.et14b,panesar.et16b,panesar.et18,sterling.et16, sterling.et17}) has until recently
been under appreciated.  

There are still however some suggestions that jets can occur when there is flux emergence in the absence of
cancelation, as argued by \cite{mulay.et16} for 6 of 20 events they studied (the remaining 14 events all
clearly included cancelation).  While we recognize that the Sun may choose to make jets in this fashion
sometimes, we also caution that for jets near active regions, it can be difficult to isolate emergence from
cancelation (\cite{sterling.et16,sterling.et17}).  Because a large majority of jets are clearly triggered by
cancelation in quiet Sun and coronal holes, where cancelation and emergence can be more easily
differentiated (\cite{panesar.et16b,panesar.et18}), one might expect cancelation to trigger almost all jets
in active regions too.  This, however, must be resolved by careful study of more jet events.   

Finally, it is now clear that some jets can reach the outer corona and make ejections in the form of narrow
CMEs or steamer-puff CMEs.  There is even the thrilling possibility that some of this material might be
observed {\it in situ} by  the Parker Solar Probe, which in turn could tell us more about the nature of
jets.

\ack
The author thanks R. L. Moore, N. K. Panesar, and D. A. Falconer for their contributions to the discussed content. This work was supported by funding from the NASA Heliophysics Guest Investigators program, and 
the MSFC Hinode project.

\section*{References}

\bibliography{ms_test}




\end{document}